\documentstyle[12pt]{article}

\begin{document}
\thispagestyle{empty}

\begin{center}
\LARGE \tt \bf{ Cosmic Rotation Axis, Birefrigence and Axions to detect Primordial  torsion fields}

\end{center}

\vspace{1cm}

\begin{center} {\large L.C. Garcia de Andrade\footnote{Departamento de
F\'{\i}sica Te\'{o}rica - Instituto de F\'{\i}sica - UERJ

Rua S\~{a}o Fco. Xavier 524, Rio de Janeiro, RJ

Maracan\~{a}, CEP:20550-003 , Brasil.

E-Mail.: garcia@dft.if.uerj.br}}
\end{center}

\vspace{1.0cm}

\begin{abstract}
Nodland Ralston (PRL,1997) investigated the cosmological anisotropy of electromagnetic fields.In this paper we show that it is possible obtain a torsion correction to Nodland-Ralston action starting from the massive Proca electrodynamics in Riemannian spacetime and performing the minimal coupling with torsion.We end up with an action which contains the Nodland Ralston action without breaking the gauge invariance.This mechanism however gives a photon a mass generated by the nonlinear torsion terms.The torsion vector is along the cosmic rotation axis and interacts with the massive photon.This method which breaks conformal invariance allow us to determine a primordial torsion of the order $10^{-29}eV$ from the well-known photon mass limits.
\end{abstract}      
\vspace{1.0cm}       
\begin{center}
\Large{PACS number(s) : 98.80 Es}
\end{center}

\newpage
\pagestyle{myheadings}
\markright{\underline{ Cosmic rotation axis and primordial torsion. }}

\section{Introduction}
\paragraph*{}

Recently some attempts have been made by Dobado and Maroto  \cite{1} to explain cosmological birefrigence phenomenon proposed by Nodland and Ralston (NR) \cite{2} on the basis of a non-Riemannian structure called Cartan torsion \cite{3}. Besides Dobado and Maroto \cite{1} tried to explain the cosmological birefrigence by primordial tiny torsion fields effectively coupled to the electromagnetic fields. Despite the fact that the NR proposal has been rebuted by some authors like Carroll et al. \cite{4} and Loredo et al. \cite{5} we could argued that the effect has not yet been confirmed but remains an interesting theoretically possibility which justifies even the investigation of corrections to this effect as proposed here. The coupling is obtained after integrating out charged fermions in a gauge invariant procedure.Although their method had the advantage of not introducing any new physics, such as massive photons, they also argued that for their procedure to be compatible with the NR ideas it would be necessary that the torsion vector field $S^{\mu}$ should be massless while the low energy effective quantum gravity
assumes that the torsion field would have a mass of the order of Planck mass.In this brief report we show that one can solve some of these problems by simply starting from the Proca massive electrodynamics in Genexral Relativistic spacetime \cite{6} and considering a minimal coupling procedure with torsion on this action which allow us to end up with the NR action with a nonlinear correction which is given by the interaction of the massive photon with the Primordial torsion field which is aligned along the Cosmic rotation axis \cite{7,8,9}.Concerning objections to the fact that we have a breaking of conformal invariance and also a breaking of the gauge invariance from the very beginning we like to mention two interesting arguments ,the first by Huang et al. \cite{10} that states that in this kind of situation is better to give a photon a mass than the graviton a mass to avoid electromagnetic catastrophes.The second interesting argument comes from Widrow and Turner \cite{11} which states that the very small limit on the photon mass \cite{12,13} does not violate any new known experiments in physics.Besides in the model presented here we end up with a cancelation of the gauge invariance breaking exactly when the photon mass is gerated by the torsion field which seems to be in agreement with electrodynamics previously proposed \cite{12,13}.Also here we obtain the interesting result that when the massive photons propagate orthogonally to the cosmic rotation axis the torsion correction dissapears and we end up with the original NR action.Most important is our assumption that the torsion axial vector lies along the direction of cosmic rotation \cite{14} axis being the root of anisotropy. It is important to stress that some authors \cite{15} considers that cosmological models in Einstein-Cartan gravity should always be anisotropic in order that the cosmological principle be valid \cite{15}.Let us star our Letter writing down the effective Proca Lagrangean
\begin{equation} 
S_{eff}=\int{d^{4}x[-\frac{1}{4}F^{2}+\frac{1}{2}{m^{2}}_{\gamma}A^{2}]}
\label{1}
\end{equation}
where $A^{2}=A_{\mu}A^{\mu}$ where $A^{\mu}$ $({\mu}=0,1,2,3)$ is the electromagnetic potential vector and $F^{2}=F_{{\mu}{\nu}}F^{{\mu}{\nu}}$ where $F_{{\mu}{\nu}}={\nabla}_{\mu}A_{\nu}-{\nabla}_{\nu}A_{\mu}$ is the electromagnetic field couplin where in principle non-minimal coupling has been used.Here ${\nabla}$ represents the Riemannian covariant derivative.Now by using the Cartan frame where the full non-Riemannian connection can be written as
\begin{equation}
{\Gamma}^{\alpha}_{{\beta}{\gamma}}={\Gamma}'^{\alpha}_{{\beta}{\gamma}}+2S^{\alpha}_{{\beta}{\gamma}}
\label{2}
\end{equation}
where S represents the torsion tensor and ${\Gamma}'$ is the Riemannian affine connection.With the minimal coupling with torsion the electromagnetic field tensor becomes
\begin{equation}
F_{{\mu}{\nu}}={F'}_{{\mu}{\nu}}+2S^{\alpha}_{{\mu}{\nu}}A_{\alpha}
\label{3}
\end{equation}
where here now $F_{{\mu},{\nu}}$ represents the full non-Riemannian electromagnetic tensor and $F^{'}_{{\mu},{\nu}}={\partial}_{\mu}A_{\nu}-{\partial}_{\nu}A_{\mu}$ represents the usual Maxwell Minkowskian electromagnetic tensor field.Substitution of expression (\ref{3}) into (\ref{1}) yields
\begin{equation} 
{S_{eff}}'=\int{d^{4}x[-\frac{1}{4}{F'}^{2}-\frac{1}{2}{F'}^{{\mu}{\nu}}S^{\lambda}_{{\mu}{\nu}}A_{\lambda}-2S^{{\mu}{\nu}}_{\lambda}A^{\lambda}S^{\beta}_{{\mu}{\nu}}A_{\beta}+\frac{1}{2}{m^{2}}_{\gamma}A^{2}]}
\label{4}
\end{equation}
The action (\ref{4}) can be expressed in the form of NR effective action 
\begin{equation}
{S_{eff}}_{NR}=\int{d^{4}x[-\frac{1}{4}{F'}^{2}+\frac{{\Lambda}^{-1}}{4}{F'}_{{\mu}{\nu}}A_{\alpha}s_{\beta}{\epsilon}^{{\mu}{\nu}{\alpha}{\beta}}]}
\label{5}
\end{equation}
where ${\epsilon}$ is the Levi-Civita totally skew-symmetric symbol, ${\Lambda}^{-1}$ is the NR constant of the order of $10^{-32} eV$ in natural units and $s_{\mu}$ is the axion vector in general associated with the cosmic rotation axis.To accomplish this task we need to cancel the Proca term which can be done by considering
\begin{equation}
m^{2}_{\gamma}={\alpha}^{2}s^{2}{\Lambda}^{-2}
\label{6}
\end{equation}
where $s^{2}=s^{\alpha}s_{\alpha}$ can be normalized.Expression (\ref{6}) implies that the massive photon must be generated by the axial torsion since the other constraint to obtain the NR action is 
\begin{equation}
S^{{\mu}{\nu}{\lambda}}=- \frac{\alpha}{\sqrt{2}}{\epsilon}^{{\mu}{\nu}{\lambda}{\beta}}s_{\beta}{\Lambda}^{-\frac{1}{2}}
\label{7}
\end{equation}
Besides we also must have the constraint
\begin{equation}
A^{\beta}s_{\beta}=0
\label{8}
\end{equation}
Unfortunatly condition (\ref{8}) is not always possible since this would imply physically that an observer in this universe would only observe photons which are ortogonal to the cosmic rotation axis and since this is in general fixed light would not be observed in all directions in the universe as is usual the case from the the well-known result of COBE-CMBR data. Another situation would be to consider that the direction of $s^{\mu}$ is not fixed in space and that Cartan torsion vector is therefore not orthogonal to the propagation of the massive photons.This is in fact possible according to a recent paper by C.Wolf \cite{16}.Let us show that it is possible to show that a conservation of a fermion (torsion) bosonic (photons) current leads to an expression for $s^{\mu}$  in terms of the gradient of an axion field which is 
in agreement with a NR result.To obtain this result let us take the interaction part of Lagrangean above as 
\begin{equation}
L{\alpha} -J^{\mu}A_{\mu}
\label{9}
\end{equation}
Comparison between expressions (\ref{4}) and (\ref{9}) yields a vector current in terms of the Maxwell field and the torsion tensor as
\begin{equation}
J^{\lambda}= \frac{1}{2}{F'}^{{\mu}{\nu}}S^{\lambda}_{{\mu}{\nu}}
\label{10}
\end{equation}
Imposing current conservation to this expression one obtains
\begin{equation}
{\partial}_{\lambda}J^{\lambda}= \frac{1}{2}[{\partial}_{\lambda}{F'}^{{\mu}{\nu}}] S^{\lambda}_{{\mu}{\nu}}+\frac{1}{2}{F'}^{{\mu}{\nu}}{\partial}_{\lambda}S^{\lambda}_{{\mu}{\nu}}
\label{11}
\end{equation}
for a totally skew symmetric torsion $S_{{\mu}{\nu}{\lambda}}=S_{[{\mu}{\nu}{\lambda}]}$ one may note that the first term on the RHS of expression (\ref{11}) vanishes dua to the second Maxwell equation ${\partial}_{[{\lambda}}F^{'}_{{\mu}{\nu}]}=0$.Besides the vanishing of the second term on the RHS due to the constraint of current conservation ${\partial}_{\lambda}J^{\lambda}=0$ yields the following condition on torsion
\begin{equation}
{\partial}_{\lambda}S^{\lambda}_{{\mu}{\nu}}=0
\label{12}
\end{equation}
But since the torsion tensor is now totally skewsymmetric it can be written in terms of the torsion vector or the direction $s^{\mu}$ of the cosmic rotation as $S^{\lambda}_{{\mu}{\nu}}={{\epsilon}^{\lambda}}_{{\mu}{\nu}{\theta}}s^{\theta}$ which from (\ref{12}) yields the final condition 
\begin{equation}
{\partial}_{[{\lambda}}s_{{\theta}]}=0
\label{13}
\end{equation} 
which is fully satisfied by the axionic NR condition $s_{\theta}= \frac{g}{8}{\partial}_{\theta}{\phi}$ where ${\phi}$ is the NR axion field.Note also that from knowledge of the limit for the photon mass allows us to determine a limit for the torsion vector of the order $|S^{\mu}|=10^{-29} eV$.This limit is by the way ten orders of mgnitude higher than the Dobado Maroto limit for the primordial torsion field.Another interesting physical consequense of this letter is that the possible determination of the plane of propagation of massive photons can be used as a astrometrical method for the determination of the Cosmic rotation axis of our universe. A recently Chern-Simons type theory proposed by Sen Gupta and his group \cite{17,18} claims to solve the birefrigence problem and proposes to indirectly detect torsion from Syncroton radiation of nebulae. In their theory the Kalb-Rammond fields play the role of torsion fields.
\section*{Acknowledgments}
\paragraph*{}
I thank C.N. Ferreira , N.O.Santos , A.Wuensche, I.Shapiro and Yuri Obukhov for helpful discussions on the subject of this letter. Financial assistance from Universidade do Estado do Rio de Janeiro(UERJ) and CNPq. is grateful acknowledged.

\end{document}